\documentclass[preprint]{vgtc}               

\ifpdf
  \pdfoutput=1\relax                   
  \usepackage{graphicx}                
  \DeclareGraphicsExtensions{.pdf,.png,.jpg,.jpeg} 
\else
  \usepackage{graphicx}                
  \DeclareGraphicsExtensions{.eps}     
\fi%

\graphicspath{{figures/}{pictures/}{images/}{./}} 

\usepackage{microtype}                 
\PassOptionsToPackage{warn}{textcomp}  
\usepackage{textcomp}                  
\usepackage{mathptmx}                  
\usepackage{times}                     
\usepackage{cite}                      
\usepackage{tabu}                      
\usepackage{booktabs}                  

\usepackage{booktabs}
\usepackage{multirow}

\onlineid{0}

\vgtccategory{Research}

\vgtcinsertpkg

\preprinttext{To appear in an IEEE VGTC sponsored conference.}

\ieeedoi{xx.xxxx/TVCG.201x.xxxxxxx}
\usepackage{caption}
\usepackage{subcaption}

\usepackage{color}
\definecolor{mscolor}{rgb}{0,0,0.7}

\title{Does the Layout Really Matter? \\A Study on Visual Model Accuracy Estimation}

\author{Nicolas Grossmann\thanks{e-mail: nicolas.grossmann@cg.tuwien.ac.at}\\ %
        \scriptsize TU Wien %
\and Jürgen Bernard\thanks{e-mail: bernard@ifi.uzh.ch}\\ %
     \scriptsize University of Zurich
\and Michael Sedlmair\thanks{e-mail: michael.sedlmair@visus.uni-stuttgart.de}\\ %
     \scriptsize University of Stuttgart
\and Manuela Waldner\thanks{e-mail: waldner@cg.tuwien.ac.at}\\ %
     \scriptsize TU Wien} %

\teaser{
  \centering
  \includegraphics[width=0.67\linewidth]{intro_2.png}
  \caption{These two different-looking plots show the results of the same image classification model. The images are paired with their predictions as indicated by colored borders. The top plot utilizes a strong grouping of the data items, while the bottom one does not. We investigated how these differences affected the users' capability to predict a model's accuracy visually.}
  \label{fig:teaser}
}

\abstract{In visual interactive labeling, users iteratively assign labels to data items until the machine model reaches an acceptable accuracy. A crucial step of this process is to inspect the model's accuracy and decide whether it is necessary to label additional elements. In scenarios with no or very little labeled data, visual inspection of the predictions is required. Similarity-preserving scatterplots created through a dimensionality reduction algorithm are a common visualization that is used in these cases. Previous studies investigated the effects of layout and image complexity on tasks like labeling. However, model evaluation has not been studied systematically. We present the results of an experiment studying the influence of image complexity and visual grouping of images on model accuracy estimation. We found that users outperform traditional automated approaches when estimating a model's accuracy. Furthermore, while the complexity of images impacts the overall performance, the layout of the items in the plot has little to no effect on estimations.
} 

\CCScatlist{
  \CCScatTwelve{Human-centered computing}{Visu\-al\-iza\-tion}{Empirical studies in visualization}{}
  \CCScatTwelve{Human-centered computing}{Visu\-al\-iza\-tion}{Visualization design and evaluation methods}{}
}

\begin{document}


\firstsection{Introduction}

\maketitle
Interactive machine learning (IML) combines the strengths of humans and algorithms and enables rapid, focused, and incremental learning cycles through which a model's training progress can be steered~\cite{amershi2014power}. Visual interactive labeling~\cite{bernard2018vial} (VIAL) is a promising strategy for users to assign one or multiple labels to one or multiple data elements and train a classification model. Color-coded similarity-preserving scatterplots (SPS) are a commonly used visualization technique~\cite{chatzimparmpas2020state} to enable users explore data characteristics~\cite{liu2017towards}, steer the labeling process~\cite{bernard2018vial}, evaluate the underlying classification model~\cite{kucher2020}, and explain misclassifications~\cite{rauber2016visualizing}. These SPS are 2D plots created through dimensionality reduction~(DR)~\cite{chatzimparmpas2020state} of the input data or activation vectors extracted from a neural network~\cite{rauber2016visualizing}.
Observable data patterns in SPS can either directly be based on structural data characteristics revealed with unsupervised DR methods, or be based on label distributions, and class predictions revealed with supervised DR methods, or both (see Figure~\ref{fig:scatter_H}).
In any case, the final layout of SPS is influenced by data-inherent properties and parameters of the DR algorithm. It has been shown that a stronger visual grouping of similar data items in the layout supports users performing labeling tasks~\cite{bernard2017comparing}. However, labeling is only one part of the VIAL process. Model evaluation, i.e., analyzing the model's performance, is the crucial next step. As especially in the early stages of the labeling process, ground truth (GT) labels for most items are unknown, automatic evaluation approaches like leave-one-out cross-validation get unstable~\cite{kohavi1995study}. Therefore, it is helpful to have humans in the loop who visually inspect the model's classification quality and decide if it is necessary to label more elements~\cite{bernard2017comparing}.

Previous studies on SPS in the IML context have focused primarily on visual interactive labeling~\cite{bernard2017comparing,chegini2020}. We complement these studies by investigating SPS for accuracy estimation of an image classification model. Layout factors, such as the grouping of classes, are considered to be crucial for SPS \cite{sips2009selecting,wang2017perception}. We, therefore, systematically investigate how the similarity-preserving layout and (predicted) class grouping influence users' ability to estimate the model's accuracy.
To this end, we conducted a crowd-sourced user study where we ask users to evaluate the results of an image classification model by estimating its accuracy through visual inspection of the predictions.

The study aims at answering the following questions:\\
    \textbf{RQ1} Can users reliably estimate a model's classification accuracy from SPS?\\
    \textbf{RQ2} How does visual image complexity affect accuracy estimation?\\
    \textbf{RQ3} How does visual grouping affect accuracy estimation? \\
    \textbf{RQ4} Does grouping facilitate model accuracy estimation for the users subjectively? \\
    
We contribute a study whose results provide the first evidence that the similarity-preserving layout and visual class grouping have little influence on users' visual model accuracy estimations.

\section{Related Work}
Recent studies focused on the effect of visualizations and the layout in SPS on the labeling process but did not take model evaluation into account. For example, Bernard~et~al. and Chegini~et~al.~\cite{bernard2017comparing,chegini2020} have shown that VIAL can outperform traditional active learning methods when the shown SPS visually separates the underlying class structures. However, they did not systematically investigate the influence of the layout. Sips~et~al.~\cite{sips2009selecting} and Wang~et~al.~\cite{wang2017perception} also focused on the beneficial aspects of perceived separability between classes in SPS. However, these effects were only shown for SPS encoding GT data, which differs from our setup where we ask users to check whether the predicted label, which is shown through colored borders, matches the image content. Work with a focus on the perceptual aspects of visual search tasks like the one by Haroz~et~al.~\cite{haroz2012capacity} or Gramazio~et~al.~\cite{gramazio2014relation} show that grouping benefits the search for outliers and the overall heterogeneity estimation of elements. Only simple color-coded plots were used for these works, whose pre-attentive features might not be applicable for VIAL. Here, users also have to interpret image content in combination with color-coding.

Model evaluation without ground truth requires (selective) instance-level inspection of images~\cite{luo2021texture}. Hoque~et~al.~\cite{hoque2013cider} used 2D image embeddings as a tool to create hierarchically grouped image layouts that allow users to more efficiently narrow down their search target. They found that grouped embeddings allow users to solve complex search tasks faster, yet not more accurately. Similar experiments were conducted by Strong~et~al.~\cite{strong2010visual}. Users preferred search interfaces where images are ordered in separate clusters based on their content. Still, they were not more accurate. Their works also showed that image complexity impacts search performance considerably. The work by Yang~et~al.~\cite{yang2020visual} shows a way to perform model evaluation by letting users evaluate the accuracy of a model by showing them the prediction results for a single image along with visual explanations and asking them whether they agree with the prediction. These works required users to find or evaluate only individual images. Our goal is for users to acquire an overview of the embedding as a whole.

\section{Experiment}
In our experiment, we investigated three aspects that can make visual model evaluation challenging in IML scenarios: (1) For the vast majority of data items, the labels are unknown. Instead of GT labels, we can only visualize predicted labels, and these predictions may be false. (2) The data-inherent features, which define the spatial proximities of items in the SPS, may not reflect the desired class structure. (3) The complexity of the data requires an instance-level inspection to evaluate the predicted labels. To systematically explore the effects of these factors, we define them as the independent variables of our experiment. For (1), we vary the \textit{accuracy} of the model predictions displayed, and for (2), we vary the grouping in the DR plot based on the data-inherent features (\textit{feature-strength}) and predicted labels (\textit{degrees of supervision}). Lastly (3), we test how the \textit{data complexity} affects the estimation results by using different kinds of image datasets, namely MNIST, Fashion-MNIST~(FMNIST)~\cite{xiao2017online}, and the animals with attributes dataset (AwA)~\cite{xian2018zero}.

To simulate different \textbf{accuracies}, we developed a greedy algorithm that computationally selects items to be labeled to achieve a dedicated target accuracy. This algorithm is based on the greedy approaches used as an upper bound in active learning~\cite{bernard2017comparing}. We adapted it to select the sample that minimizes the difference between the model accuracy and the set target accuracy. As an additional constraint, the target classes in the training data are balanced. For each dataset, we trained a KNN classifier to approximate \textit{low} (50\%), \textit{medium} (75\%) and \textit{high} (100\%) target accuracy, by labeling 20 out of 400 samples. The actual model accuracies are shown in Figure \ref{fig:results}.

\begin{figure}
     \centering
     \includegraphics[width=0.4\textwidth]{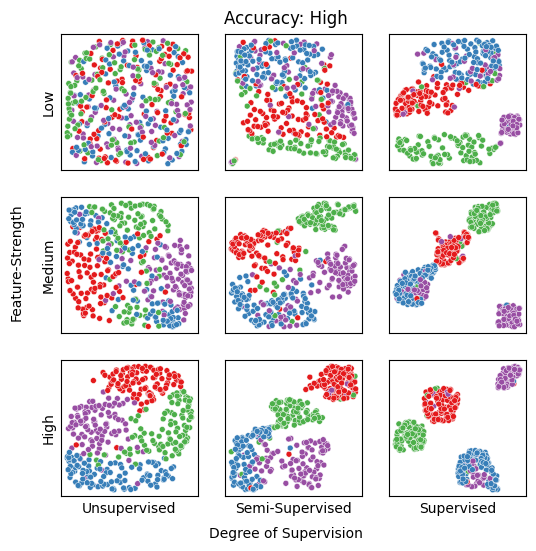}
     \caption{The same data elements across all nine between-subject conditions (\textit{feature-strength} $\times$ \textit{degree of supervision}). Here, dots are the images, and color encodes the GT labels (not predicted labels). 
     }
     \label{fig:scatter_H}
\end{figure}

DR methods try to preserve data-inherent patterns in their lower-dimensional representation. Ideally, those patterns follow the expected class distribution and group elements of similar classes in the SPS. This grouping may or may not be true in reality. To emulate less ideal cases, we apply a Principal Component Analysis (PCA) to the image feature vectors before applying the DR to vary the \textbf{feature-strength} systematically. With PCA, we obtain the principal components ranked by their variance. By selecting PCs with lower eigenvalues as feature vectors, we can artificially decrease the data-inherent grouping of classes. For \textit{high} feature-strength, we selected the original data (Figure~\ref{fig:scatter_H}); \textit{medium} features use 75\% of data-inherent variance; for the \textit{low} feature condition, we only chose 50\%.

To enhance grouping in the layout, we vary the \textbf{degree of supervision} during DR. For this, we chose UMAP~\cite{mcinnes2018umap} as it allows us to control the influence of global and local patterns. The unsupervised layout, which is commonly used~\cite{chatzimparmpas2020state}, relies entirely on the image features and therefore reveals data-inherent patterns. In contrast, the supervised layout aims to maximize class separation. In our case, this creates a visual grouping based on predicted labels. 
We chose to create three layouts for each input data configuration: unsupervised, fully supervised, and semi-supervised with a partial label set (Figure~\ref{fig:scatter_H}). To avoid image overlaps, we apply quantile-based normalization to the data.

To investigate how datasets of varying visual \textbf{image complexity} affect the users' performance, we chose two different kinds of image datasets to be tested in our study. Fashion-MNIST~\cite{xiao2017online} is a commonly used dataset in the machine learning community. It contains easily identifiable images of clothing items, easily identifiable even on an overview level. On the other hand, animals with attributes~\cite{xian2018zero} is a dataset that has complex images of animals in various poses, that requires users to look more closely.

\subsection{Task}
By scaling up the approach by Yang~et~al.~\cite{yang2020visual}, we derived our model evaluation task. We showed users 400 images that should be associated with four classes. Each image has a colored border representing the model prediction (Figure~\ref{fig:teaser}). The class color association is shown in a legend. Based on this information, users should then estimate the percentage of correctly predicted image classes, i.e., the model accuracy. The result could be set using a slider from 0\% to 100\%. We provide users with standard interaction methods like zooming and panning. Users are shown a one-minute timer as a guideline to self-check how long they have been working on a task. Users were instructed to finish within one minute, but the system did not enforce the time limit. This temporal guidance was chosen based on a pilot study with the goal to prohibit users from counting all individual items independently of the layout.

\subsection{Design and Procedure}
We used a mixed design, with \textit{degree of supervision} and \textit{feature-strength} as between-subject factors and \textit{accuracy} and \textit{image complexity} as within-subject factors. We employed all layout variations as between-subjects factors as qualitative feedback of twelve pilot users indicated that users employ different principal task solving strategies based on the amount and quality of visual grouping. To avoid adverse learning effects for users who are shown both well separated and mixed up patterns, we kept the grouping defined by the feature-strength and supervision constant within users. Each user was then shown in random order all three GT accuracies of the FMNIST and AwA datasets. In addition, the MNIST dataset was used as a tutorial example as the first task. Here, the users are also informed on how well they estimated the accuracy.

Before the study, participants were asked basic demographic questions like age, gender, and proficiency with visualization and machine learning. Here we also conducted a simple color-blindness test with Ishihara plates. This test was necessary as color vision was required to differentiate the colored borders. After all tasks, users were asked if they found the visualization helpful and were confident in their choices using 5-point Likert scales.

\subsection{Apparatus}
The study was hosted on Amazon Mechanical Turk (MTurk). Users who signed up for the study were re-directed to our server. The visualizations were created on the fly using JavaScript and D3. Although our layout creation process creates spread-out images across the whole area, we also use a force-based algorithm to push overlapping images apart, ensuring that each image can be inspected individually. A unique ID was assigned to each worker, and their responses, task completion time, and motions were logged. The server assigned each participant one of 27 configurations.

To assure the quality of the responses, we used an attention task after the first half of the tasks. Users were shown a SPS of 400 images from four classes out of the MNIST dataset with ground truth labels. Users who guessed anything below $80\%$ accuracy on this task were excluded. If users either failed the attention task or the color-blindness test initially, they were instructed to return the HIT. The task was then re-assigned to a new user, and the previous user received no penalty on their MTurk metrics. 

We also inspected the results manually to check if users provided reasonable responses, i.e., not submitting the same response for every task. From a pilot study with nine users, we determined a maximal completion time of 15 minutes. Assuming a fair wage of 12 USD an hour, we paid each user that completed the study 3 USD.


\subsection{Participants}
We collected the data of 74 users, at least eight for each configuration. The number was decided following other crowdsourced and visual search studies~\cite{hoque2013cider,strong2010visual}. As an added benefit, many MTurk workers perform labeling tasks regularly. Therefore, we required participants to have completed 10.000 HITs with an average approval rate above $95\%$ that are commonly used to ensure data quality~\cite{peer2014reputation,robinson2019tapped}. Users were also screened for common forms of color-blindness using seven Ishihara plates. A challenge for visual MTurk studies is to account for different screen resolutions and input devices. Therefor we performed a pre-screening, where only users of non-mobile devices were allowed to proceed.

Out of 74 users, two users were excluded due to being outliers as they either took longer than 70 seconds on average time or skipped over tasks. Their tasks were then re-assigned. The mean age of users was $36.6$, and $39.4\%$ were female. Only $18.3\%$ of users stated above-average knowledge about data visualization and $31.0\%$ with regards to machine learning.

\section{Results \& Discussion}
In total we, obtained 432 measurements ($72 \textrm{ users} \times 3 \textrm{ accuracies} \times 2 \textrm{ image complexities}$). We tested the dependent variables for normality and applied parametric or non-parametric tests as appropriate to answer our four research questions. As we only have up to eight data points per between-subject conditions, there is a risk of Type II errors. We, therefore, also report the effect sizes. Details of our analysis are in the supplemental document.

\begin{figure*}
    \centering
    \begin{subfigure}[b]{0.625\textwidth}
        \centering
        \includegraphics[width=\textwidth]{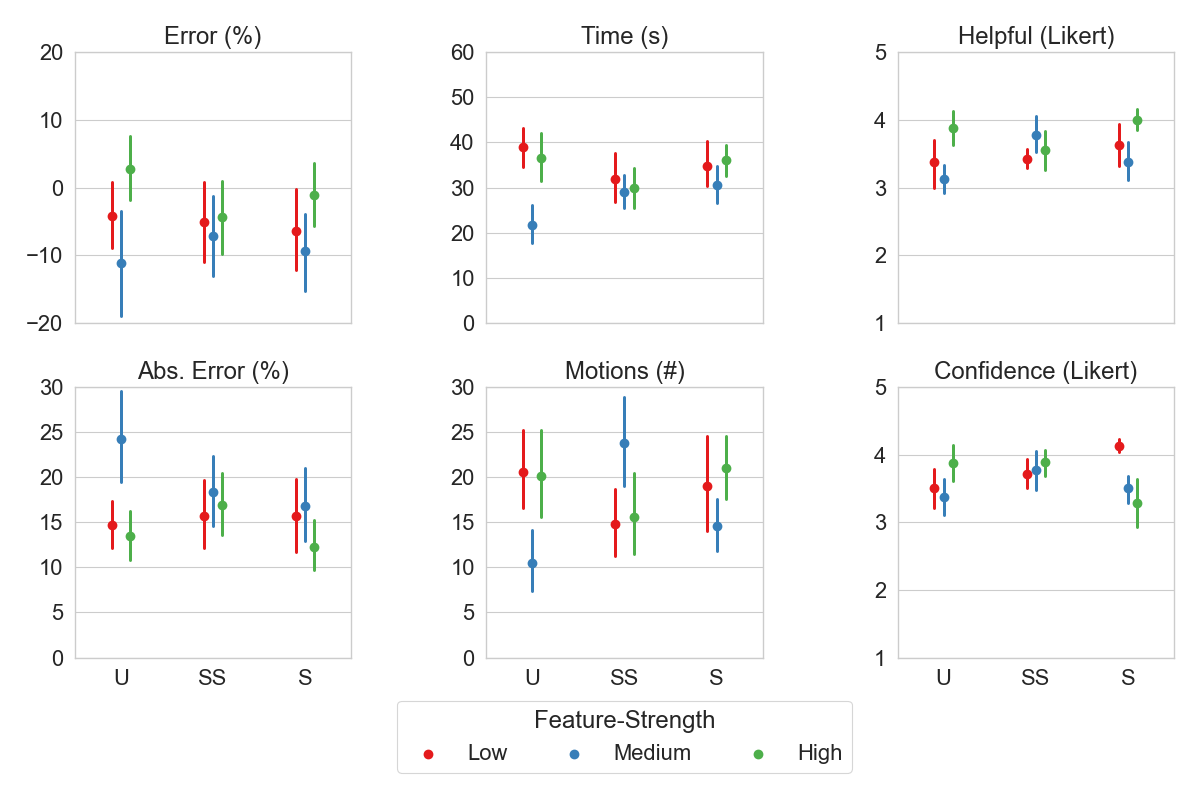}
        \caption{Aggregated scores across all between-subject conditions (CI 95\%). Degree of Supervision: Unsupervised (U), Semi-Supervised (SS), Supervised (S). We see no significant differences in the dependent variables across the different grouping factors.}
        \label{fig:conditions}
    \end{subfigure}
    \hfill
    \begin{subfigure}[b]{0.325\textwidth}
        \centering
        \begin{subfigure}[b]{\textwidth}
            \centering
            \includegraphics[width=0.95\textwidth]{results.png}
            \caption{Estimated accuracies for all three accuracy levels and both datasets. (black lines: actual accuracies)}
            \label{fig:results}
        \end{subfigure}
        \vspace{2ex}
        \begin{subfigure}[b]{0.8\textwidth}
            \centering
            \includegraphics[width=\textwidth]{reg.png}
            \caption{Negative correlation between the absolute error and the time spent on the task.}
            \label{fig:regression}
        \end{subfigure}
    \end{subfigure}
    \label{fig:overview_results}
    \caption{Overview of the study results.}
\end{figure*}

To answer whether users can estimate model accuracies from SPS in general (\textbf{RQ1}), we performed a Friedman test to see if users provided distinct estimates for low, medium, and high model accuracies. The results show a large and significant difference between users' accuracy estimations dependent on the model's accuracy (FMNIST: $\chi^2(2, N = 72) = 42.5; p < 0.001$, AwA: $\chi^2(2, N = 72) = 40.9; p < 0.001$). Compared to traditional cross-validation (Figure~\ref{fig:results}), humans are more accurate in their estimations. Yet, users tend to overestimate low accuracies (FMNIST: $3.85\%$, Awa: $7.57\%$)  and underestimated medium (FMNIST: $-9.83\%$, Awa: $-6.01\%$) and high accuracies (FMNIST: $-10.9\%$, Awa: $-16.0\%$). In summary, these results show that \emph{users can visually estimate model accuracies fairly reliably, while cross-validation leads to unstable accuracy estimations}.

To assess how image complexity affects accuracy estimation (\textbf{RQ2}), we performed a paired samples t-test. The average time to judge the model accuracy, based on 400 images, is $32.1$ seconds. There exists a strong, significant difference between the two datasets (FMNIST: $29.1$, AwA: $35.1$; $t(71) = -5.727; p < 0.001; d = -0.675$). The number of motions follows the same patterns: The average is $17.8$ motions (zooms and pans), with a large, significant difference between the datasets (FMNIST: $13.6$, AwA: $22.1$; $t(71) = -7.330; p < 0.001; d = -0.864$). The higher image complexity of the second dataset leads to a significantly higher completion time and number of interactions, but the estimation results are still comparable (FMNIST: $-5.83\%$, AwA $-4.80\%$; $t(71) = -0.467; p = 0.642; d = -0.055$).


There are moderately strong correlations between estimation error, completion time, and the number of motions (Figure~\ref{fig:regression}). Time shows a strong positive correlation with the number of motions ($r(70) = 0.742;p < 0.001$) and a moderate with the signed error ($r(70) = 0.453; p < 0.001$). Time also has a moderate, negative correlation with the absolute estimation error ($r(70) = -0.622; p < 0.001$).
These correlations indicate that user behavior has a substantial effect on the estimation error. Quick users viewed the data on a higher zoom level and only focused on few images, while slower users spent more time zooming in on details and panned over large plot regions. The more closely users inspected the results, the better their estimations. To compensate for the different user behaviors when analyzing the between-subjects factors, we used time and error, respectively, as covariates for the following analysis steps.

We then investigated the influence of layout (i.e., feature-strength and degree of supervision) on estimation error, completion time, and amount of interactivity (\textbf{RQ3}) using an ANCOVA. Feature-strength has a medium effect on the estimation error that fails to reach significance ($F(2, 62) = 2.538; p = 0.087; \eta^2_p = 0.076$). As shown in Figure~\ref{fig:conditions}, users tend to underestimate the error less for higher feature-strength (L: $-6.25\%$, M: $-7.70\%$, H: $-1.59\%$). The effect of feature-strength on time ($F(2, 62) = 1.245; p = 0.295; \eta^2_p = 0.039$) and motions is very small and insignificant ($F(2, 62) = 0.228; p = 0.797; \eta^2_p = 0.007$). The degree of supervision has a very small and insignificant effect on error ($F(2, 62) = 0.213; p = 0.809; \eta^2_p = 0.007$), time ($F(2, 62) = 0.461; p = 0.633; \eta^2_p = 0.015$), and motions ($F(2, 62) = 0.252; p = 0.778; \eta^2_p = 0.008$). There is also no relevant interaction effect between feature-strength and supervision on the error ($F(4, 62) = 0.313; p = 0.869; \eta^2_p = 0.020$), time ($F(4, 62) = 0.522; p = 0.720; \eta^2_p = 0.033$), or motions ($F(4, 62) = 1.548; p = 0.199; \eta^2_p = 0.091$).

These results suggest that grouping has little to no influence on users' estimations. The layout variants we tested were highly diverse, ranging from nearly randomly arranged images in the case of unsupervised DR and low feature-strength up to tightly packed homogeneous clusters in high feature-strength and full supervision (Figure~\ref{fig:teaser}). Our results indicated no benefit of separated clusters for the task of model evaluation, neither for the completion time nor the estimation error. These results are in contrast to previous studies~\cite{bernard2017comparing,haroz2012capacity} where visual grouping facilitated task performance. One contributing factor could be that this task could not be solved pre-attentively. It required the user to scan the images, even in the case of simple images. Scanning the images group-wise did not facilitate this task to any considerable extent. We assume that it was sufficient to selectively zoom into any part of the image canvas to estimate the amount of incorrectly classified images within this region. With such a zoomed-in view, the overall layout probably played a minor role in the accuracy estimation. These results indicate that the task is more similar to visual search tasks like the ones described by Hoque~et~al.~\cite{hoque2013cider}.

Overall, most users found the shown SPS helpful (H; $56.9\%$) and were confident (C; $72.2\%$) in their accuracy estimations (\textbf{RQ4}). A Kruskal-Wallis test showed no statistically significant preferences for plots depending on degrees of supervision (H: $\chi^2(2, N = 72) = 0.362; p = 0.834$; C: $\chi^2(2, N = 72) = 0.707; p = 0.702$) or feature-strength(H: $\chi^2(2, N = 72) = 1.914; p = 0.384$; C: $\chi^2(2, N = 72) = 0.373; p = 0.692$).

Summarized, users appropriately judged their ability to predict the accuracy and found SPS primarily helpful. However, we found no indications that the grouped layouts were perceived as more helpful or provided users more confidence in their predictions.

\section{Study Limitations \& Future Work}
The \textbf{participants} of our study were crowd-sourced workers, as they represent non-expert users that very likely come in touch with data labeling tasks. However, non-expert users may not have the same motivation and capabilities as expert users to infer knowledge from the grouped layouts. A closer investigation of business analysts or data scientists actively developing models would require a different recruitment strategy and study procedure but might yield interesting results.

The \textbf{task} of model evaluation is only one sub-task of a larger visual IML workflow, which was studied in isolation here. Future studies could investigate the effects of grouping and image complexity on other tasks like outlier detection or searching for specific sub-groups in the data. Another option would be to test other visual encodings than SPS for visual model evaluation. If the layout is not crucial for this task, maybe simpler representations are even more effective.

The \textbf{size} of the dataset in terms of shown images and classses was either larger~\cite{hoque2013cider} or similar~\cite{bernard2017comparing} to previous studies but may still be limited compared to real-world IML scenarios. Nevertheless, our results suggest that selectively looking at a subset of elements may be sufficient to estimate model accuracy. Further studies could analyze the scalability of model evaluation and labeling tasks both in terms of items displayed and the number classes. 


\section{Conclusion}
SPS are widely used to visualize model performance~\cite{chatzimparmpas2020state}. Our experiment shows that users can estimate the model accuracy reliably in an IML scenario, where the SPS shows images with their predicted labels. Coming back to the question proposed in the title, we can say that our experiment indicates that the DR-driven layout has a negligible influence on the task performance and users' satisfaction when used for visual model evaluation, requiring instance-based inspection. Our results point to that the effective use of SPS relies on the task at hand, and their use should be critically examined.


\acknowledgments{
This work was funded by the Digital Society Initiative (DSI) at University of Zurich and the German Research Foundation (DFG), Project-ID 251654672 – TRR 161.}

\makeatletter
\interlinepenalty=10000
\bibliographystyle{abbrv-doi}

\bibliography{references}
\makeatother
\end{document}